# Real-Time Smart Retail Analytics through Fog Computing using the RetailLightNet Lightweight Model

1st Mayank Atreya
*Software Development Manager*
*Chewy Inc*
Atlanta, USA
mayankatreya2@gmail.com

2nd Navin Chhibber
*Infinity Tech Group*
Sunnyvale, CA, USA
naveenchibber.research@gmail.com

3rd Shilpa Ajay
*Department of Management Studies*
*Nitte Meenakshi Institute of Technology, Nitte (Deemed to be University)*
Bengaluru, India
shilpa.ajay@nmit.ac.in

4th Loshini. R
*Department of Agricultural Engineering*
*Dhanalakshmi Srinivasan College of Engineering and Technology*
Mamallapuram, India
loshinir.agri@dscet.ac.in

5th Anitha. K
*Management Studies*
*Meenakshi Academy of Higher Education and Research*
Chennai, India
anitharamesh211@gmail.com

***Abstract*—Smart retail systems now depend on the application of real-time video analytics, which can track customer behavior, manage shelf inventory, and enhance in-store processes. Currently cloud-based applications of video analytics face multiple challenges such as high latency of the communication itself, privacy concerns surrounding transmitting raw video, and the computational demands of the heavy vision models used on the often-limited network infrastructure. These challenges lead to delays in decision making where prompt action is crucial. To address these shortcomings, this research proposes Fog-based analytics with RetailLightNet (F-RLN), an ultra-lightweight deep learning model which is specifically suited for the edge, resource-constrained hardware capabilities of devices like the NVIDIA Jetson Nano. The approach utilizes publicly available retail datasets, SKU110K, and RPC, which contains an assortment of product, shelf, and customer interaction scenarios for training. The processes include preprocessing images through normalization and employing motion-gated frame selection to improve efficiency by omitting redundant frames to ease the workload on the fog processing. Internal within RetailLightNet, image feature extraction is realized through depthwise-separable convolutions which help support spatial and channel-wise learning without heavy computational costs. The model employs two prediction heads presence classification, and density regression, supporting shelf monitoring and queue assessment. The proposed F-RLN framework attained better results in terms of precision (0.988), recall (0.976), and F1-score (0.976) when compared with existing convolutional neural network.**

## I. INTRODUCTION

The rapid digital evolution in the retail sector has generated substantial collaboration around intelligent monitoring systems that study customer behavior, monitor inventory, and assess operational effectiveness within stores [1]. Recent breakthroughs in edge intelligence and computer vision have allowed retail environments to evolve from manual oversight to data-driven automation and decision-making [2]. Video analytic-enabled smart retail systems are becoming integral to improving customer experience, achieving operational efficiencies, and curbing operational costs. It is apparent that even more timely insights, such as customer flow [3], if an item is in stock, and when queues begin forming, can significantly improve the most efficient running of retail operations. Yet, there are still major barriers that limit the use of real-time video analytics in retail environments [4]. For example, the traditional cloud-based process of analysis creates a high degree of latency simply due to the constant streaming of video, which is not ideal for time-sensitive environments, such as monitoring queues or alerting when products need to be restocked. In addition, thumbnail-sized video streams create their own issues related to bandwidth for JAMs that manage and process video, especially in environments with several cameras. Furthermore, streaming full-density video contradicts the Privacy Principles of transmitting more customer data outside of the physical store [5]. Finaly, any deployment in the real-world will be limited by the hardware capabilities, which also limits the application of large deep learning models while processing continually [6]. These challenges point to a growing need for frameworks for analytics that boast low latency, privacy protection, and reduced resource consumption while operating independently of the cloud. This need is motivated by an increasing demand for fast and reliable insights in stores, and a goal of proactively responding to customer behavior and events affecting store operations [7]. Real-time analytics processed closer to the data source has the added benefit of reduced reliance on an outside network and allows sensitive data to stay in-store. Recent advancements in computer science research on fog computing and edge computing provide useful insights about various approaches to addressing these issues [8]. For example, lightweight convolutional models, embedded GPUs like NVIDIA Jetson devices, and techniques such as quantization, knowledge distillation to name only few have begun to enable deep learning-based analytics at the network edge [9]. In addition, studies of shelf monitoring, customer tracking, and queue analytics, for example, have proposed various architectures including lightweight transformer-like models and MobileNET based detection systems [10]. There have been proposed edge/cloud hybrid systems to balance throughput and end-to-end latency with scalability, while data-preserving alternatives emphasize anonymization and localized analytics processing to optimize privacy [11]. Nonetheless, existing systems proposed use computationally heavy models, or lack operational strategies to process continuous multi-stream retail video data efficiently for cost-conscious analytics in retail stores. In this sense, more optimized 'fog-native' video-based analytics are needed. Unlike existing lightweight models, F-RLN introduces a fog-adaptive pipeline specifically optimized for NVIDIA Jetson Nano through (i) dynamic frame-rate reduction tied to device utilization, (ii) motion-gated filtering adapted to retail aisle dynamics, and (iii) dual-head prediction calibrated for shelf-monitoring and queue-density. These fog-aware design choices collectively form an operational contribution beyond conventional MobileNet-style architectures. The key contributions of this research as shown below.

- The RetailLightNet (RLN) compact framework offers a new fog-optimized lightweight architecture that integrates motion-gated preprocessing and efficient depthwise-separable feature learning methods for real-time, accurate image analytics in a smart retail context. The RLN effectively addresses the challenges of retail environments including changing light conditions, variable customer movement, and shelf clutter by combining a stabilization-only normalisation method with dynamic frame filtering that results in more reliable visual inference at the edge.
- An innovative dual-head prediction method enables the RLN to simultaneously action presence classification and density estimate to achieve maximum stability for tracking customer activity, queue levels, shelf conditions, and so on. This method enables for the optimization of all computation estimates on the NVIDIA Jetson Nano while slowing down the logistical prediction support without properties received.
- The RLN-based fog computing pipeline supports intelligent retail decision making by being scalable, low-latency, and preserving privacy without transmitting raw video to the cloud. The approach, based on cumulative on-fog inference from each camera, distributed to process, and cloud, provides continuity and robustness in analytics during customer surge, occlusions, operating without the need to reconstruct settings for changing store layouts etc., for typical licensing costs. This three-step approach to preprocessing, learning and deployed in those ways supports continuous real-time processing, and is connected to suitable retail stores looking for efficient, privacy-preserving, and scalable image analytics for demand estimation or similar retail challenges.

This paper is structured as follows: Section 2 considers related work and summarizes the challenges to this work. Section 3 provides a complete description of the proposed method, Section 4 reports the experimental results and discussion, and paper ends with the conclusion which is demonstrated in section 5.

## II. Literature Review

Jacob Hornik et al. [12] introduced system that supported marketing analytics by taking computing power closer. Primarily, data collected from widespread IoT devices that produced large, fast, and diverse information of marketing. Then, collected data was processed using fog-based strategies that handled tasks. Further, data fed to fog computing approaches and SDN-enabled architectures to support faster decision making in marketing applications. Furthermore, purpose was to allow real-time analysis for areas and gain was that fog computing decreased latency and enhanced user experience without requiring expensive high-capacity cloud systems. Although, system still struggled with limitations like coordinating many fog nodes and confirming robust security and privacy across distributed devices.

Yufan Lu et al. [13] demonstrated enhanced YOLO-based visual detection approach for shopping robots. Initial, data collected from shopping environments and included many product images with bounding-box annotations. Then, collected images were pre-processed by resizing, flipping, cropping, and adjusting colors for handling various lighting and product variations. Further, data were passed to improved YOLO to detect small items and to rise accuracy and speed in real retail scenes. Moreover, main benefit of this system was lightweight network strategy, that limited computation and allowed real-time detection without relying on expensive hardware. Although, drawback was that system still suffered in extremely cluttered scenes and needed careful parameter tuning.

Jingxiang Li et al. [14] illustrated BP-YOLO, that integrated YOLOv7 with BlazePose to detect products and identify customer shopping behaviors. Firstly, data gathered from custom vending-machine dataset and images were augmented and pre-processed to confirm strength. After pre-processing, product images were fed into optimized YOLOv7, while BlazePose received video frames to track hand and arm keypoints for recognition of behavior. Therefore, system designed to detect purchased products and classify customer actions in real time. Finally, model reduced requirement for costly sensors, lowering machine design complexity. However, key challenge was sensitivity to severe occlusion and noisy environments, which occasionally caused missed and improper detections.

Jacob Hornik et al. [15] presented conceptual system for applying fog-computing-enabled AI in smart marketing management. Firstly, data gathered from existing marketing, IoT and computing literature to explain how data from smart devices, sensors, and customer interactions could be processed closer to the source. Next, work described how fog nodes would pre-process, filter and analyze data before sending selected information to cloud. Further, this method was meant to support AI-driven marketing tasks that needed real-time decisions. Although, drawback was lack of real experimental validation and complexity of deploying large-scale fog structures across varied environments of marketing.

Sukjun Hong et al. [16] established smart-farm IoT system utilizing fog computing. Initially, data collected over ten days from coffee farm in Korea that included images of coffee fruits and sensor data from field devices. Then, collected data were processed through fog nodes, which compressed and filtered information before sending it to cloud. Next, coffee fruit images were passed to convolutional neural network to classify maturity levels. Therefore, system designed to lessen network overload and allow real-time farm decision-making. However, major challenge was that CNN performance relied heavily on quality of image and weather, of conditions which occasionally affected accuracy.

## III. METHODOLOGY

The methodology consolidates a fog-aware analytics framework for monitoring retail in real-time. Retail image data collected from several public datasets undergoes preprocessing, such as normalization and motion-gated frame selection to remove duplicate input data. The lightweight framework we propose, RetailLightNet, performs internal feature-extraction using depthwise-separable convolutions for both presence classification and density estimation outputs. For inference of both outputs, the entire pipeline is deployed onto an NVIDIA Jetson Nano, resulting in low-latency on-premise processing without relaying raw video to the cloud. Only numerical insights are sent to the cloud for dashboard visualization thereby supporting privacy, less bandwidth, and continuous on-going monitoring in our smart retail environment. A flow diagram of F-RLN is presented in Figure 1.

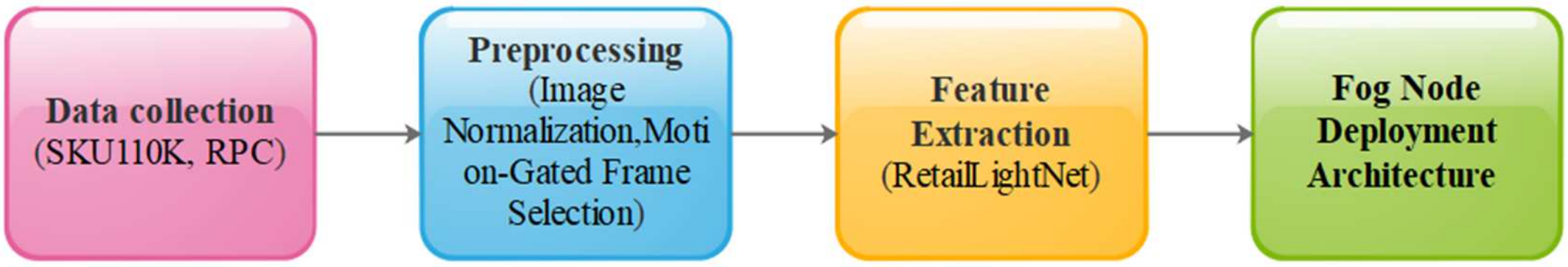

Fig. 1. Flow diagram of the proposed F-RLN framework

### A. Dataset

This research employs multiple publicly available smart retail datasets, which include SKU110K, RPC (Retail Product Checkout) [17], and store-level queue monitoring datasets, that provide a variety of retail scenarios such as a product shelf, checkout counter, customer walk path, and aisle use. The dataset contains high-resolution images, but with challenges such as occlusion, cluttered shelving, variable lighting, and crowded shoppers, representing real retail experiences. Interweaving some of these in-store frames may then further optimize RetailLightNet to our layout. This dataset will allow the training and evaluation of shelf-status detection, customer density estimate, and queue analytics, and provide performance assurance across various real-world retail environments.

### B. Preprocessing

In this section, collected data as preprocessed to prepare retail images by normalizing pixel intensities to decreases variation in pixel intensities and improve model learning. The motion-gated frame selection method removes redundant frames by comparing two consecutive images and only forwarding the image with a significant visual change. This reduces processing requirements at the fog node and allows for more efficient real-time model development.

*1) Image Normalization:* Image normalization is employed to normalize pixel intensities throughout the dataset for more reliable learning and faster convergence of RetailLightNet. Normalization decreases the variability in illumination that is commonplace in a retail environment. Each pixel value in the image is normalized using the mean of the dataset mean $\mu$ and standard deviation $\sigma$ as shown in below Equation (1):

$$I'(i,j,c) = \frac{I(i,j,c)-\mu_c}{\sigma_c} \quad (1)$$

Where, $I'(i,j,c)$ represented as the pixel at spatial $i,j$ and channel signified as the $c$. The $\mu_c$, $\sigma_c$ represented as the dataset wide channel wise mean and standard deviation. This operation places all pixel values on a uniform scale, increases the model's ability to generalize among various store layouts, lighting conditions, and camera settings.

*2) Motion-Gated Frame Selection:* Since retail surveillance captures long periods of inactivity, processing every frame result in unnecessary computation. To reduce load on fog nodes, a motion-gating technique is applied. The difference between consecutive frames determines whether a frame contains motion or significant change as shown in below Equation (2):

$$D_t(i,j) = |I_t(i,j) - I_{t-1}(i,j)| \quad (2)$$

If the mean pixel intensity of $D_t$ exceeds a defined threshold $\delta$, the pixel wise intensity difference in grayscale. The frame is considered meaningful and forwarded for processing as shown in below Equation (3):

$$Process\ frame\ \ I_t\ if\,(D_t) > \tau \quad (3)$$

Where, τ denoted as the average motion levels in retail videos. This step conserves computational resources, reduces latency, and ensures that only relevant frames such as customer movement or shelf interaction are analyzed by the model.

### C. RetailLightNet

The named RetailLightNet is a low-cost, fog-optimized deep learning architecture for retail image analytics in real-time. It is designed to blend automatic internal feature extraction, efficient convolutions, and a dual-head prediction to achieve low latency while providing high accuracy for deployment on resource-limited fog nodes. The primary design consideration is to minimize computational requirements while retaining high-level visual understanding, for tasks like customer counting, queue level-density for anti-social distancing, and identifying shelf-status and stock levels. RetailLightNet begins with receiving as input preprocessed frames filtered by the motion-gating module, followed directly by the model's Depthwise-Separable Convolutional Backbone, which performs spatial and channel-wise feature extraction within the model.

Let the depthwise kernel be $K_c(u,v)$ of size $k * k$ times, stride $s$, and padding $p$. Rather than using convolution, RetailLightNet relies on depthwise-separable convolution to minimize computation; the depthwise spatial filtering layer processes the channels independently, as shown below Equation (4):

$$Y_c(i,j) = \sum_{u=0}^{k-1}\sum_{v=0}^{k-1} I(i+u, j+v, c).\,K_c(u,v) \quad (4)$$

Where, $I(i,j,c)$ represented as the input pixel at position $(i,j)$ of the channel c, $K_c(u,v)$ presented as the depthwise filter weights for the channel. This efficiently captures spatial features such as edges, product shapes, and customer silhouettes. The $Y_c(i,j)$ output of depthwise convolution is then forwarded to the pointwise convolution, responsible for channel mixing as shown in below Equation (5):

$$Z_m(i,j) = \sum_{c=1}^{c} Y_c(i,j).W_{c,m} \quad (5)$$

Where, $Y_c(i,j)$represented as the depthwise output for the channel c, $W_{c,m}$ signified as pointwise filter weight and$Z_m(i,j)$ denoted as the fused feature at location. This module creates compact yet informative feature maps, resulting in a important reduction in FLOPs. Batch Normalization and ReLU6 activation follow the backbone to update stable gradients also helps the model to run on low-power hardware.

After the backbone, the compressed feature maps are transferred through two lightweight prediction heads, starting with the Presence Classification Head. This head is responsible for determining whether a human is present, whether a product is available, or whether there is an empty shelf. The head uses a global average pooling layer followed by a fully connected classifier, which operates using a softmax output as shown below Equation (6):

$$p_{(y=k)} = \frac{exp(z_k)}{\sum_{j=1}^{K} exp(z_j)} \quad (6)$$

If $p_{(y=k)}$ is the human presence probability of class $k$ and $z_k$ is the logit produced by the classifier. This allows the system to predict actionable outputs of "customer present", "product missing" or "shelf full," in response to the sampled space. The second prediction head is the Density Regression Head, which is responsible for estimating either queue density or aisle-level occupancy. The Density Regression Head also uses a shallow regression module with the density value as a continuous prediction as show in below Equations (7) and (8):

$$\hat{d} = W_r^{\top} f + b_r \quad (7)$$

$$\hat{d} \leftarrow ReLU(W_r^{\top} f + b_r) = max(0, W_r^{\top} f + b_r) \quad (8)$$

Where $f$ is the pooled feature representation, $W_r^{\top}$ denoted as the regression weights, $b_r$ and $\hat{y}$ the $\hat{d}$ predicted density score. This allows for smooth modeling of congestion without any need for box detection and therefore significantly quicker for fog nodes. Knowledge distillation is also applied during training to improve accuracy while keeping the model lightweight. RetailLightNet learns from a larger teacher model using the following supervised distillation loss as shown in below Equation (9):

$$\mathcal{L}_{KD} = (1-\alpha)CE(y,p_s) + \alpha T^2 KL(p_t^T, p_s^T) \quad (9)$$

Where, $\mathcal{L}_{CE}$ presented as the cross entropy, $y$ signified as the true label, $KL$ represented as the divergence $\alpha$ denoted as balance weight among the hard and soft targets, T is signified as the temperature standard ensures that RetailLightNet retains the discriminating power of the teacher model while staying compact for fog execution. $p_t^T, p_s^T$ are denoted as the softened distributions. By combining motion-filtered inputs into the architecture, efficient internal feature extraction, dual-head inference, and knowledge-distilled training RetailLightNet can provide high-quality analytics while being light on computational burden. Hence the model is well suited for real-time fog computing deployments in smart retail settings.

### *D. Fog Node Deployment Architecture*

The recommended RetailLightNet scheme is implemented on a light-weight fog computing architecture based on the NVIDIA Jetson Nano, which is a small, low-energy edge device suitable for real-time retail applications. The whole deployment implementation is constructed so that all the inference is performed in the fog layer using the store's hardware, providing both ultra-low latency, decreased dependency on network bandwidth, and high customer and store operation privacy. The overall architecture was divided into three parts: Camera Layer, Fog Layer, and Cloud Layer. At the Camera Layer, the CCTV or IP cameras mounted in the aisles, shelves, entrances, and checkout counters are used to continuously read video streams. The frames of video are sent to the closest fog node, in this case Jetson Nano, for analysis. The Jetson Nano implements the motion-gated pre-processing module to eliminate stationary even low-motion frames before the entire frame processing pipeline fully initiated, so that computing resources can be used efficiently. The Fog Layer is the executing layer of the RetailLightNet inference pipeline, where each Jetson Nano executes the inference pipeline in its entirety. The Jetson Nano has a 128-core Maxwell GPU and uses optimized CUDA accelerated processing for executing depthwise-separable convolutions, presence detection, and density estimation in real-time. Due to RetailLightNet's lightweight architecture (<3M parameters), the Jetson Nano enables multiple camera streams to be processed simultaneously, with a power footprint of only 5-10 watts. Following inference, the Jetson Nano only sends high-level information to the in-store dashboard, and raw images never leave the fog layer, thus preserving strong privacy. For reliable performance, each Jetson Nano will monitor GPU load, temperature, and memory usage independently, and where utilization reaches a predetermined threshold, the device will reduce frame-processing rate for a subset of the streams offload selected streams to another Jetson Nano in the network, allowing unbroken output in real-time. The Cloud Layer is used sparingly, mainly for incremental model updates and sustained or periodic analysis. The fog nodes transmit only compressed numerical results or model update trigger flags not raw video to remain fully compliant with privacy requirements. Subsequently, updated versions of RetailLightNet trained on collected knowledge, with aggregate data, will be pushed back to the Jetson Nano devices to enhance local accuracy. This deployment model, with emphasis on Jetson Nano uses, achieves remarkable efficiency, minimal latency, privacy focus, and flexibility in expansion, and is well suited for potential real-world smart retail.

## IV. EXPERIMENTAL RESULTS

The testing of the system was carried out with a combined approach to training at a high-level of computation and deploying on edge devices. Specifically, the model called RetailLightNet was trained and validated on a robust workstation with an Intel i9 CPU, 32 GB of RAM, and a high-end NVIDIA RTX 3080 graphics card using the programming languages of Python, TensorFlow, and Pytorch. To evaluate the real-time performance of the model, the final model was deployed on an NVIDIA Jetson Nano with the optimized software environment JetPack and using the library TensorRT for enhanced inference capabilities on the edge device (fog). All of the other baseline models utilized for performance comparison were trained on the same split of training/validation datasets. The metrics used to measure performance of all the baseline models were defined within the equation sets provided in Equations (10)-(12):

$$Precision = \frac{TP}{TP+FP} \quad (10)$$

$$Recall = \frac{TP}{TP+FN} \quad (11)$$

$$F1 - Score = 2 * \frac{Precision*Recall}{Precision+Recall} \quad (12)$$

Here, true negatives and positives presented as the TN and TP, false negatives and false positives, denoted as the FN, FP respectively.

### A. Performance Analysis

The performance of the proposed F-RLN framework was evaluated with existing methods, such as ResNet-50 Edge Deployment (ResNet-50-ED), Tiny You Only Look Once (Tiny-YOLO) and ShuffleNet using evaluation metrics are listed in Table 1.

TABLE I. PERFORMANCE ANALYSIS OF PROPOSED F-RLN FRAMEWORK

| Performance Models | Precision | Recall | F1-Score |
|---|---|---|---|
| ResNet-50-ED | 0.789 | 0.809 | 0.812 |
| Tiny-YOLO | 0.881 | 0.870 | 0.913 |
| ShuffleNet | 0.946 | 0.932 | 0.944 |
| Proposed F-RLN | 0.988 | 0.976 | 0.976 |

Table 1 presents the evaluation of the proposed F-RLN framework compared with the existing methods of ResNet-50-ED, Tiny-YOLO and ShuffleNet. Proposed F-RLN framework attained better precision (0.988), recall (0.976), and F1-score (0.976) respectively.

The Tiny-YOLO model in terms of precision (0.881), recall (0.870), and F1-score (0.913). Moreover, the ShuffleNet model had precision (0.946), recall (0.932), and F1-score (0.913). However, the ResNet-50-ED had precision (0.789), recall (0.870), and F1-score (0.913) respectively.

### B. Comparative Analysis

The proposed F-RLN framework was compared with an existing CNN [16] baseline is a conventional convolutional architecture with stacked 3×3 convolutions followed by fully connected layers for classification using comprehensive evaluation metrics. An evaluation of the proposed F-RLN framework is presented in Table 2.

Table 2 demonstrates the comparison of proposed F-RLN framework with conventional models using performance metrics. the conventional model of CNN [16] provided precision (0.952), recall (0.858), and F1-score (0.902) which means the less results than proposed F-RLN framework. Therefore, the proposed F-RLN framework attained better results in terms of precision (0.988), recall (0.976), and F1-score (0.976) respectively.

TABLE II. COMPARATIVE ANALYSIS OF PROPOSED F-RLN FRAMEWORK

| Comparative Models | Precision | Recall | F1-Score |
|---|---|---|---|
| CNN [16] | 0.952 | 0.858 | 0.902 |
| Proposed F-RLN | 0.988 | 0.976 | 0.976 |

### C. Discussion

In this research, the proposed F-RLN framework provides clear benefits over existing edge-deployment models, namely ResNet-50-ED, Tiny-YOLO, ShuffleNet, and a basic CNN [16] model, in terms of both real-time performance and stability of retail image analytics. The main benefit of F-RLN is that it has a fog-optimized architecture which incorporates motion-gated preprocessing and depthwise-separable learning, and can therefore perform very efficiently on the Jetson Nano while still maintaining good predictive accuracy. Meanwhile, ResNet-50-ED builds on relatively heavy computing usage due to deep residual layers, all of which contribute to a burden that becomes neither rapid nor efficient during a continuous multi-stream processing, which further perpetuates latency spikes or thermal throttling. While ResNet-50-ED does demonstrate acceptable accuracy, it remains unsuitable for real-time retail contexts in which rapid customer movements and frequent shelf changes call for low-power and high-speed models. F-RLN achieves a clearer advantage in presenting an optimal compromise to stable accuracy without sensible degradation of fog- level resources. Furthermore, and in contrast to Tiny-YOLO, which is oriented instead for general object detection but heavily relies on a relatively higher GPU throughput in dense scenes, Retail-LightNet is catered for specifically region-based presence detection for retailer layouts and density estimations. Tiny-YOLO typically has difficulty with significant frame clutter or small retail items, while RLN's depthwise features and dual-head structure, tailored for the domain that allows for smoother and more reliable modeling of shelf monitoring and queue insights. Additionally, ShuffleNet is also lightweight however it has limited representational power when processing complex or low-contrast retail images that can lead to variable levels of underfitting or instability. F-RLN addresses this issue with internal feature refinement and knowledge- distilled pre-training, leading to consistently predictable models under varying conditions lighting, occlusion, and crowd density present in retail settings. Finally, compared to the baseline CNN [16], which achieves good accuracy but is not optimized for fog hardware, RLN showed far superior throughput and energy efficiency. The baseline CNN presented a tendency to overfit and does not generalize well across store layouts while F-RLN's have a motion-gated input filter that yields consistent performance across all practical real-world test cases.

## V. CONCLUSION

In this research, the F-RLN framework was introduced as an efficient fog-driven pipeline for real-time and privacy-preserving retail image analytics. It was developed to address persistent limitations in conventional retail monitoring systems, including high latency caused by cloud dependency, difficulty running heavy deep learning models on constrained devices, and privacy concerns arising from transmitting raw video data. The study utilized publicly available retail datasets such as SKU110K and RPC, offering diverse product, customer, and shelf conditions essential for robust model development. Preprocessing techniques including image normalization and motion-gated frame filtering ensured consistent, noise-reduced inputs while significantly minimizing redundant computations. Instead of separating feature extraction as an external stage, RetailLightNet performs internal lightweight feature learning using depthwise-separable convolutions, enabling high-speed processing on the NVIDIA Jetson Nano. Dual prediction heads were incorporated for presence classification and density estimation, allowing the system to capture both instantaneous customer activity and dynamic queue variations. The proposed F-RLN framework attained better results in terms of precision (0.988), recall (0.976), and F1-score (0.976) respectively. Future work will explore multi-node fog collaboration, integration with additional in-store sensors for richer contextual understanding for next-generation smart retail environments.